\documentstyle[aps,prd,eqsecnum,epsf,twocolumn]{revtex}
\newcommand{\be}{\begin{equation}}
\newcommand{\ee}{\end{equation}}
\newcommand{\bea}{\begin{eqnarray}}
\newcommand{\eea}{\end{eqnarray}}

\title{Construction of Cosmic String Induced Temperature
 Anisotropy Maps with CMBFAST and Statistical Analysis}
\author{N.~Simatos\dag ~and L.~Perivolaropoulos\ddag }
\address{\dag Department of Physics, University of Crete,\\
GR--71003 Heraklion, Crete, Greece\\
\ddag Institute of Nuclear Physics, \\
National Centre for Scientific Research ``Demokritos N.C.S.R.'',\\
Athens, Greece\\
e--mail: {\tt simatos@physics.uoc.gr,\ leandros@mail.demokritos.gr}}

\date{\today}
\draft

\begin{document}

\maketitle

\begin{abstract}
We use the publicly available code CMBFAST as modified by L.~Pogosian
and T.~Vachaspati to simulate the effects of wiggly cosmic strings on
the Cosmic Microwave Background (CMB). Using the modified CMBFAST code which takes into account
vector modes and models wiggly cosmic strings by the one scale
model, we go beyond the angular power spectrum to construct CMB
temperature maps with resolution of a few degrees. The statistics
of these maps are then studied using conventional and recently
proposed statistical tests optimized for the detection of hidden
temperature discontinuities induced by the Gott-Kaiser-Stebbins effect. 
We show however, that these realistic maps can not be distinguished in 
a statistically significant way from purely Gaussian maps with identical power spectrum.
\end{abstract}


\section{Introduction}
A promising method of identifying the origin of the large--scale
structure of the universe, is the statistical analysis of the CMB
fluctuations
\nocite{gott90,moes94,mag95,f-m97,l-a-m99}[1-5]. What one
hopes to get out of this is the possible distinction between the
two major classes of physical theories of structure formation,
namely theories of inflation \cite{lyth99} and topological defects \cite{vs94}. An important
differentiating  feature of these two classes of theories is the
statistics of primordial fluctuations. Temperature anisotropies
$\frac{\Delta T}{T}$ of the CMB, if originating from quantum
fluctuations in an inflationary context, should obey Gaussian
statistics. However, the confidence by which the CMB anisotropies
were shown to follow that statistic  is
reduced \cite{kogut96}, taking in mind the work done by \cite{per-sim98,f-m-g98}
using COBE--DMR maps. This could favor topological defect
theories.

Cosmic strings seeding structure formation
\nocite{p-s-t97,a-b-r97,a-b-r99,b-r-a98,c-h-m99}[11-15] consist a 
promising case of these theories. However, despite the significant 
progress made recently 
\nocite{a-b-r97,a-b-r99,tur-pen-sel98,sel-pen-tur97a,sel-pen-tur97b}[12,13,16-18] 
in understanding the effects of cosmic strings on the CMB, the 
uncertainty of the derived predictions remains significant 
\cite{deruelle99,dur-sak97}. The main source of this uncertainty 
is that there is no simple way to characterize the network of 
strings. In order to bypass this difficulty various studies have 
attempted to model the string network by incorporating simplifying 
assumptions. Such simplifications have made possible the 
realization of cosmological string network simulations 
\nocite{a-c-d-k-s97,a-c-s-s-v96,ben-bouch90}[21-23] that attempt 
to model evolution in a very wide range of scales. They have also 
made possible the construction of semi-analytical models that 
attempt to capture features missed by the inherent limitations of 
full scale numerical simulations. For example, early work 
\cite{per93,per95} based on the latter approach had revealed 
important features of the cosmic string induced CMB fluctuations 
that were missed by detailed numerical simulations. In particular 
it was shown\cite{per95}, using a simple model of string network 
evolution, that the{\em\ wiggles} (small scale structure) of long 
cosmic strings tend to amplify the height of the Doppler peak of 
the CMB angular spectrum of string induced fluctuations. 

The main reason for this amplification may be attributed to  the 
two main factors that determine the string induced fluctuations. 
These are, first the{\em\ Integrated Sachs--Wolfe effect} (long 
strings present between last scattering and the present time 
interacting gravitationally directly with the photons) for scales 
larger than 2 degrees and second, for scales about 1-2 degrees, 
interactions of the photons with the string perturbed plasma 
performing  acoustic oscillations on the last scattering surface 
(LSS) excited by long strings.  It may be shown 
\nocite{per95,lev-vac99,vv91}[25-27] that the latter effect leads 
to temperature fluctuations ${\Delta T}\over T$ that are amplified 
compared to the fluctuations induced by the former effect. The 
amplification factor $\lambda > 1$ is due to the fact that the 
wiggles do not interact gravitationally directly with the photons, 
while they do interact with the plasma \cite{per95,vv91} on the 
LSS leading indirectly to additional fluctuations on the CMB 
through the last scattering of photons on the fast moving plasma 
electrons. Therefore this amplification can only affect the last 
scattering horizon scale, i.e. a scale of about 1-2 degrees. The 
factor $\lambda$ can be expressed as \cite{per95,vv91} 
\begin{equation}
\lambda=(1+{{(1-{T\over \mu})}\over {2<(v_s \gamma_s)^2 >}})
\end{equation}
where T is the tension of the wiggly long string  which decreases
with the wiggliness  and is estimated by simulations to be
$T\simeq 0.7\mu$ where $\mu$ is the mass per unit length of the
string. Also, $v_s$ is the string velocity and $\gamma_s$ the
corresponding Lorentz factor.

This result of Doppler peak amplification due to cosmic string
wiggles has recently been verified by the numerical modeling
developed by Pogosian and Vachaspati \cite{lev-vac99}. These
authors used the line of sight integration approach of the 
publicly available code CMBFAST \cite{zal-sel96} combined with 
the{\em\ one--scale} model \cite{kibble85,kibble86} for wiggly 
cosmic string evolution to derive the angular CMB power spectrum 
and the matter power spectrum. One of their main results was that 
small scale structure (wiggles) of long strings tends to increase 
the height of the Doppler peak of the CMB spectrum thus confirming 
the expectations of Ref. \cite{per95} and improving the agreement 
of the predicted angular spectrum with observations. 

In the present work we use the approach of Pogosian and Vachaspati 
and modify their publicly available code based on CMBFAST to go 
beyond the CMB spectrum and construct realistic maps of CMB 
fluctuations produced by wiggly strings. Our goal is to apply 
statistical tests on several realizations of these maps and 
attempt to identify their non-Gaussian features.  Some of the 
tests we applied \cite{per-prd98} (which were specially designed 
for this purpose) were able to identify non-Gaussian features on 
simple CMB maps \cite{per-sim98}, constructed by the superposition 
of a small temperature discontinuity (produced by a single long 
string via the{\em\ Gott-Kaiser-Stebbins effect} 
\cite{gott85,kais-steb84}) on a Gaussian background. 
 In the present study however,
we show that these same tests are not able to identify the string
induced non-Gaussianity of more realistic maps on the scales
considered. This is due to the fact that the superposition of
several line-like temperature  discontinuities tends to wash out 
the effects of each individual discontinuity and reduces the 
effectiveness of the proposed tests. Therefore, the detection of 
the small non-Gaussianity induced by cosmic strings on a few 
degrees scale remains a challenging open issue. 

\section{Construction of the maps}

The construction of the CMB temperature anisotropy maps on which
the statistical tests were performed, was made possible by two
major contributions: the wiggly string {\em\ one--scale} model as
the cosmic string network model and the use of the line of sight
integration approach.

The cosmic string network model that we used is the one found in
detail in \cite{lev-vac99}. Briefly, the network is approximated
\nocite{a-b-r97,a-b-r99,b-r-a98}[12-14] by
a collection of uncorrelated straight string segments which are
assumed to be produced at some early epoch, with positions drawn
randomly from a uniform distribution in space, moving  with random
uncorrelated velocities, their directions drawn from a uniform
distribution on a two sphere. This model has the merit of being
relatively simple and amenable to modifications that seem to be
indicated by direct simulations \cite{ben-bouch90,ben-bouch-steb88}.

Small--scale structure,{\em\ wiggles,} were incorporated in order to make the model more realistic, since lattice simulations of string formation and evolution \cite{ben-bouch90} suggest that strings are not straight.
However, a distant observer would not be able to discern this small--scale structure, but see instead a smooth string with effective mass per unit length
$\tilde{U}$ and tension $\tilde{T}$. The wiggly string is heavier and slower
than the ordinary  Nambu-Goto straight string (whose equation of state is:
$\mu_{0}=T$, where $\mu_{0}$ is the mass per unit length).
The equation of state of a wiggly string (averaged over the small--scale structure) is \cite{carter90,vilenkin90}:
\bea
\nonumber
\tilde{U}\tilde{T}=\mu_{0}^{2},\\
\tilde{U}=\alpha \mu_{0},\ \tilde{T}=\mu_{0}/\alpha
\eea
where $\alpha$ is the ``wiggliness'' parameter, estimated \cite{ben-bouch90}
in the radiation and matter eras to be $\alpha_{r}\simeq1.9$ and $\alpha_{m}\simeq1.5$
respectively. The expected evolution of the wiggliness parameter is fitted by \cite{ben-bouch90,lev-vac99}:
\be
\alpha(\tau)=1+\frac{(\alpha_{r}-1)a}{\tau\dot{a}}
\ee
where $a$ is the scale factor.

The parameters of the segments: length, velocity and wiggliness 
are modeled using the {\em\ one--scale} model 
\cite{kibble85,kibble86}. The evolution of the two competing 
processes of string stretching on the one hand (due to cosmic 
expansion), and the chopping off of loops and their subsequent 
decay on the other (due to long string reconnection), are 
described by \cite{mar-shel96}: \bea 
\frac{dl}{d\tau}=\frac{\dot{a}}{a} lv^{2}+\frac{1}{2} \tilde{c}v\\ 
\frac{dv}{d\tau}=(1-v^{2})(\frac{\tilde{k}}{l}-2 
\frac{\dot{a}}{a}v) \eea where $l$ is the comoving correlation 
length, $v$ is the $rms$ string velocity, $\tilde{c}$ is the loop 
chopping efficiency and $\tilde{k}$ is the effective curvature of 
the strings \cite{a-b-r99}. At every subsequent epoch, a certain 
fraction of the number of segments decays in a way that preserves 
scaling. 

The Fourier transform of the energy--momentum tensor for an individual string segment $m$ is:
\bea
\Theta_{00}^{m}=\frac{\mu\alpha}{\sqrt{1-v^2}}\frac{\sin{(k\hat{X}'_{3}l/2)}}{k\hat{X}'_{3}l/2}\cos{({\bf{k}}\cdot{\bf{x_{0}}}+k\hat{\dot{X}}_{3}v\tau),}\\
\Theta_{ij}^{m}=\left[v^{2}\hat{\dot{X}}_{i}\hat{\dot{X}_{j}}-\frac{(1-v^{2})}{\alpha^{2}}\hat{X'}_{i}\hat{X'}_{j}\right]\Theta_{00},
\eea
while $\Theta_{0i}^{m}$ can be found from the relation: $\nabla^{\mu}
\Theta_{\mu\nu}^{m}=0$. ($X^{\mu}(\sigma,\tau)$ are the coordinates of the segments: $X^{0}=\tau,\hspace{2mm}{\bf{X}}={\bf{x}_{0}}+\sigma\hat{{\bf{X'}}}+v\tau\hat{{\bf{\dot{X}}}}$, where ${\bf{x}_{0}}$ is the random location of the center of mass, $\hat{{\bf{X'}}}$ and $\hat{{\bf{\dot{X}}}}$ are
 unit vectors along and perpendicular to the string which are
randomly oriented, also satisfying $\hat{{\bf{X}'}}\cdot\hat{{\bf{\dot{X}}}}=0$
and $\sigma$ is the coordinate along the string).

A consolidation of all string segments that decay at the same epoch
into one with the appropriate statistical weight is used, as was
suggested in \cite{a-b-r97}, so that the number of segments can be dealt with computationally.
 The total energy--momentum tensor of the network is the sum over the contributions of the consolidated string segments \cite{lev-vac99}:
\be
\Theta_{\mu\nu}({\bf{k}},\tau)=\sum^{N_{0}}_{m=1}\Theta_{\mu\nu}^{m}({\bf{k}},\tau) T^{off}(\tau,\tau_{m}),
\ee
where $N_{0}$ is the initial number of segments, and $T^{off}$ is a smooth function that turns off the $m^{th}$ 
segment by time $\tau_{m}$ \cite{a-b-r97,a-b-r99}. 
$\Theta_{\mu\nu}({\bf{k}},\tau)$ will be incorporated into the sources: $S(k,\tau)$
(cf eqs.(\ref{integr-approach},\ref{dl})).

Our main goal is to compute the CMB anisotropy seeded by cosmic
strings:
\be
\frac{\Delta T}{T}(\hat{{\bf{n}}})\equiv\frac{\Delta T}{T}({\bf{x}}=0,
\hat{{\bf{n}}},\tau_{0})=\sum_{l,m}a_{lm}Y_{lm}(\hat{{\bf{n}}}) 
\label{delta}
\ee 
(${\bf{x}}$ is the position of the observer and $\hat{{\bf{n}}}$ is the line of sight 
direction). The code\footnote[1]{http://theory4.phys.cwru.edu/\~\ 
\hspace{-1.3mm}levon} of \cite{lev-vac99} (based on CMBFAST 
\footnote[2]{http://www.sns.ias.edu/\~\ \hspace{-1.3mm}matiasz/CMBFAST/cmbfast\\.html}) which we modified 
and used, uses the line of sight integration approach 
\cite{zal-sel96} to compute, among other things, the angular power 
spectrum $C_l$: 
\be
C_{l}=\frac{1}{2l+1}\sum^{l}_{m=-l}<a_{lm}^{*}a_{lm}> \ee 
originating from temperature perturbations seeded by cosmic 
strings by incorporating  the sources into $\Theta_{\mu\nu}$ in 
the way described below. 

\vspace*{2mm} The Fourier transform 
$\Delta({\bf{k}},\hat{{\bf{n}}},\tau_{0})$ of $\frac{\Delta 
T}{T}(\hat{{\bf{n}}})$ depends only on the angle $q$ between the 
two vectors $(q\equiv\hat{{\bf{k}}}\cdot\hat{{\bf{n}}})$ and can 
be expanded in multipole moments $\Delta_{l}({\bf{k}},\tau_{0})$ 
\cite{ma-berts95}: 
\be
\Delta({\bf{k}},\hat{{\bf{n}}},\tau_{0})=\sum^{\infty}_{l=0}(-i)^{l}
(2l+1)\Delta_{l}({\bf{k}},\tau_{0})P_{l}(\hat{{\bf{k}}}\cdot\hat{{\bf{n}}})
\label{Leg-expan}
\ee
\vspace*{2mm}
\noindent
further, $\Delta_{l}({\bf{k}},\tau_{0})$ is decomposed into \cite{ma-berts95}:
\vspace*{2mm}

\be
\Delta_{l}({\bf{k}},\tau_{0}) \equiv \xi({\bf{k}}) \Delta_{l}(k,
\tau_{0})
\ee
\vspace*{2mm}

However, according to the line of sight integration approach \cite{zal-sel96,sel-measur97}:
\vspace*{2mm}

\be
\Delta(k,q,\tau_{0})=\int_{0}^{\tau_{0}} e^{i k q(\tau-\tau_{0})} 
S(k,\tau)d\tau \label{integr-approach} \ee \vspace*{2mm} \noindent 
where the term $S(k,\tau)$ contains the contributions from all the 
sources, as well as $\Theta_{\mu\nu}$ from the string network 
\cite{zal-sel96}. Using the relation: 
\be
e^{i k q(\tau-\tau_{0})}=\sum_{l}(-i)^{l}(2l+1)
j_{l}[k(\tau_{0}-\tau)] P_{l}(q) \ee we get \cite{zal-sel96}: 
\be
\Delta_{l}(k,\tau_{0})=\int_{0}^{\tau_{0}}S(k,\tau)
j_{l}[k(\tau_{0}-\tau)] d\tau
\label{dl}
\ee

In order to get the $a_{lm}$ coefficients needed to construct the
maps, we have to assume arbitrary direction for $\hat{{\bf{k}}}$, or
equivalently use the more general relation \footnote[3]{L.D. Landau,  in {\em Quantum Mechanics}, Chap.~V,~\S~34, p.\ 113.}:
\be
e^{i{\bf{k}}\cdot{\bf{r}}}=4\pi\sum_{l=0}^{\infty}\sum_{m=-l}^{l}
i^{l}j_{l}(kr)Y^{*}_{lm}(\hat{{\bf{k}}})Y_{lm}(\hat{{\bf{r}}}) 
\ee 
in (\ref{Leg-expan},\ref{integr-approach}). Otherwise we would be 
able to compute only the $a_{l0}$ terms as we shall see next. Then 
(\ref{dl}) still holds. The $a_{lm}$s read \cite{ma-berts95}: 
\bea 
\nonumber a_{lm}&=&\int d\Omega Y^{*}_{lm}(\hat{{\bf{n}}})\frac{\Delta 
T}{T}(\hat{{\bf{n}}}) =4\pi(-i)^{l} \int d^{3}{\bf{k}} 
Y^{*}_{lm}(\hat{{\bf{k}}})\Delta_{l}({\bf{k}}, \tau_{0})\\ &=&4\pi(-i)^{l} 
\int d^{3}{\bf{k}} Y^{*}_{lm}(\hat{{\bf{k}}})\xi({\bf{k}})\Delta_{l} (k,\tau_{0}) 
\eea where $\Delta_{l}(k,\tau_{0})$ is given in (\ref{dl}) (had we 
not assumed the arbitrariness of $\hat{{\bf{k}}}$, we would only get: 
$a_{lm}= (-i)^{l}\sqrt{4\pi(2l+1)} \int d^{3}{\bf{k}} \xi({\bf{k}}) 
\Delta_{l} (k,\tau_{0}) \delta_{m0}=a_{l0})$. 
\vspace*{2mm}

Since the linearized Einstein equations may be decomposed into
scalar, vector and tensor parts (S, V, T) \cite{crittenden-turok95},
we have three distinct contributions to the sources (e.g. S: density
perturbations, V: cosmic strings, T: gravitational radiation,
 \cite{a-b-r99,tur-pen-sel98}). So, $(\Delta,
\Delta_{l})$ are actually: $(\Delta^{I},\Delta^{I}_{l})$, where
$I\equiv
S,V,T$. Then:
\be
a_{lm}=a^{S}_{lm}+a^{V}_{lm}+a^{T}_{lm},
\ee
while $C_{l}=C^{S}_{l}+C^{V}_{l}+C^{T}_{l}$. Following the
normalization used in the code, the $a_{lm}$s are
defined as:
\bea
\label{alms}
\nonumber
a^{S}_{lm}=\sqrt{\frac{4}{\pi}} \int d^{3}{\bf{k}} Y^{*}_{lm}(\hat{{\bf{k}}})
\Delta^{S}_{l}({\bf{k}},\tau_{0})\\
a^{V}_{lm}=\sqrt{\frac{4}{\pi}l(l+1)} \int d^{3}{\bf{k}} Y^{*}_{lm}(\hat{{\bf{k}}})
\Delta^{V}_{l}({\bf{k}},\tau_{0})\\
\nonumber
a^{T}_{lm}=\sqrt{\frac{4}{\pi}\frac{(l+2)!}{(l-2)!}}
\int d^{3}{\bf{k}} Y^{*}_{lm}(\hat{{\bf{k}}})\Delta^{T}_{l}({\bf{k}},\tau_{0})
\eea
where: $\Delta^{I}_{l}({\bf{k}},\tau_{0})\equiv \xi({\bf{k}})
\Delta^{I}_{l}(k,\tau_{0})$. In order to implement this scheme, we
had to extend the code of \cite{lev-vac99} to provide for the amplitude of the initial perturbations
$\xi({\bf{k}})$, which must be such that \cite{ma-berts95,sel-measur97,sel-zal-allsky97}:
\be
<\xi^{*}({\bf{k}})\ \xi({\bf{k'}})>=P_{\xi}(k)\ \delta(
{\bf{k}}-{\bf{k'}})
\ee
where $P_{\xi}(k)$ is their power spectrum.

As in \cite{lev-vac99}, we chose the initial power spectrum to be that
of{\em\ white--noise}, i.e. $P_{\xi}(k)= $ constant. Consequently,
$\xi({\bf{k}})$  was chosen to be proportional to a uniformly random number:
$\xi_{{\bf{k}}}\ \epsilon[-1,1]$. Since: $<\xi_{{\bf{k}}}\xi_{{\bf{k'}}}>=\frac{1}{3}\delta_{{\bf{kk'}}}$, each $a^{I}_{lm}$ was multiplied by a factor of $\sqrt{3}$ when the ensemble average $<a^{I*}_{lm}a^{I}_{lm}>$ was taken in order to compute and verify the $C^{I}_{l}$s
\footnote[4]{The $C^{I}_{l}$s are: 
$C^{S}_{l}=\frac{4}{\pi}\int k^{2}dk|\Delta^{S}_{l}
(k,\tau_{0})|^{2},\
C^{V}_{l}=\frac{4}{\pi}l(l+1)\int k^{2}dk|\Delta^{V}_{l}(k,\tau_{0})|^{2}$ and \
$C^{T}_{l}=\frac{4}{\pi}\frac{(l+2)!}{(l-2)!}\int k^{2}dk|\Delta^{T}_{l}(k,\tau_{0})|^{2}$.}.

Applying: $\int d^{3}{\bf{k'}}<\xi^{*}({\bf{k}})\ \xi({\bf{k'}})>=P_{\xi}(k)$ to our discrete case we get:
$\sum_{{\bf{k}}} \Delta V_{{\bf{k}}} <\xi_{{\bf{k}}}\xi_{{\bf{k'}}}>=\frac{1}{3}\sum_{{\bf{k}}}
\Delta V_{{\bf{k}}}\delta_{{\bf{kk'}}}=\frac{1}{3}\Delta V_{{\bf{k}}}$.
This leads to the definition of $\xi({\bf{k}})$ as:
\bea
\nonumber
\xi({\bf{k}}) \equiv \frac{\xi_{{\bf{k}}}}{\sqrt{\Delta
V_{{\bf{k}}}}} ,\hspace{4mm} \xi_{{\bf{k}}}: random\ \epsilon[-1,1]\ ,\\
\Delta V_{{\bf{k}}} \equiv k^{2} dk\ d\Omega(\hat{{\bf{k}}}) 
\eea 
Having chosen this particular realization for $\xi({\bf{k}})$, we can proceed to
the construction of the temperature anisotropy maps using eqs.(\ref{delta},\ref{alms}).

\section{Results}

\begin{figure}
\centerline{\epsfxsize = 0.85\hsize \epsfbox{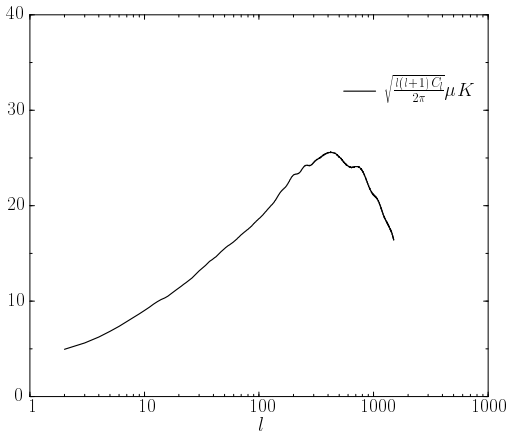}}
\caption[FIG.1]{Angular power spectrum of a $l=1500$ map.}
\label{fig1}
\end{figure}

We chose to construct maps with $l=40$, so that the pixel size is
comparable to that of the COBE maps, although our method enables
us to proceed even up to much higher $l$s (e.g. $l\simeq1500$),
with greater cost in computational time though. The angular power
spectrum of such a map is depicted in Fig.\ref{fig1} in agreement
with Ref. \cite{lev-vac99}. 

\begin{figure}[bp]
 \centerline{\epsfxsize = 0.86\hsize \epsfbox{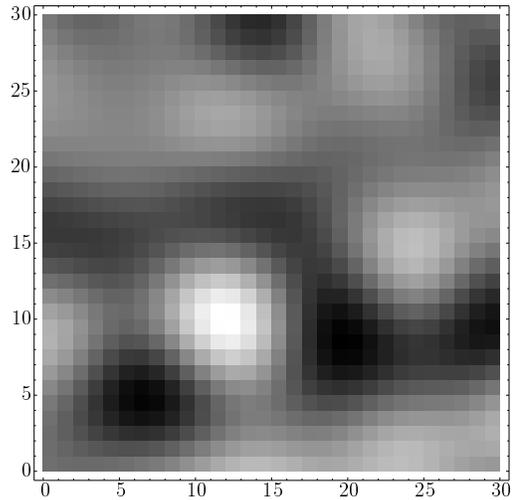}}
\caption[FIG.2]{A $30^{\circ}\times30^{\circ}$ patch of one of the 
constructed standardized temperature anisotropy maps. (Lighter 
colors correspond to higher values of $\frac{\Delta T}{T}$).} 
\label{fig2} 
\end{figure}

A typical example of such a map is depicted in Fig.\ref{fig2}, where
a $30^{\circ}\times30^{\circ}$ patch of a standardized temperature anisotropy
map is shown.

We have applied statistical tests on the above maps in an attempt 
to identify possible non-Gaussian features. Such statistical tests 
\cite{per-prd98} have recently been proposed for the detection of 
the particular type of non--Gaussianity induced by coherent 
discontinuities (edges) hidden in CMB maps. The main advantage of 
these statistical tests is that they focus on the large--scale 
coherent properties of CMB maps and are therefore effective even 
in cases of low resolution maps provided that the area covered is 
large. The statistics calculated for maps of CMB anisotropies 
seeded by cosmic strings which we have constructed are the 
skewness, kurtosis and the recently devised MSD, SMD 
\cite{per-prd98} which are optimized for the detection of coherent 
discontinuities in 1-D and 2-D pixel maps. It should be pointed 
out that the skewness and kurtosis statistics have 
proven\cite{per-prd98} to be notably unsuitable to detect the 
subtle non-Gaussianity in CMB maps. However they have been used as 
a benchmark\cite{per-prd98} to test the effectiveness of more 
complex statistics like the SMD and the MSD specially designed for 
the detection of specific types of non-gaussian signatures. 

In addition to the statistics discussed here, there are a number 
of other more complex statistics\cite{com-stat} (minkowski 
functionals and  bispectrum and higher order cumulant estimates in 
harmonic space) that may be applied to CMB maps to check for 
non-Gaussianity or topological signatures of defects. Those 
statistics which are particularly effective on high resolution CMB 
maps have been discussed extensively in the literature and their 
application on string maps produced by CMBFAST can be an 
interesting extension of the present work. Here we focus on the 
MSD and SMD statistics which are specially 
designed\cite{per-sim98} to detect the specific signature of large 
scale coherent discontinuities on CMB maps of low resolution.  

Assuming a $30^{\circ}\times30^{\circ}$ (in pixels) standardized 
temperature map $T_{ij}$($i,j=1,\ldots,30$), the skewness $s$ and 
kurtosis $k$ are defined as : 
\begin{eqnarray}
s=<T^{3}>\equiv \sum_{i,j} T_{ij}^{3}/(30)^{2},\\ k=<T^{4}>\equiv
\sum_{i,j} T_{ij}^{4}/(30)^{2}.
\end{eqnarray}
These are conventional statistics and their values for Gaussian
maps with uncorrelated pixels are $s=0, k=3$. To define the
statistics MSD and SMD we consider a partition of the CMB maps in
two parts separated by a random curve. In this study we have
considered straight lines as well as right angles. Let $\bar{k}$
denote the set of parameters that define the partition line and
let $\bar{T_{u}}$ and $\bar{T_{l}}$ be the mean temperatures of
the two parts of the map (the indices ``u'' and ``l'' stand for
``upper'' and ``lower'' parts). The statistical variable
$Y_{\bar{k}}$ is defined as :
\be
Y_{\bar{k}} \equiv |\bar{T_{u}}-\bar{T_{l}}|. \ee The statistics
MSD and SMD are defined as:
\be
MSD \equiv \frac{1}{N} \sum_{\bar{k}}Y_{\bar{k}}, \ee
\be
SMD \equiv \max(Y_{\bar{k}}), \ee where $N$ is the total number of
partitions. Both MSD and SMD approach asymptotic values for
large-$N$. Consider now a Gaussian pattern of temperature
fluctuations with a small coherent temperature discontinuity
defined by a partition $\bar{k_{0}}$ superposed on the map with
coherence scale comparable to the size of the pattern. In
\cite{per-prd98} it was shown that the presence of such a
discontinuity can be detected much more efficiently by the
statistics MSD and SMD than by the skewness and kurtosis. A
physically motivated mechanism which can lead to the production of
a coherent discontinuity on CMB maps is the presence of a moving
long cosmic string in our horizon via the fore-mentioned{\em\ 
Gott-Kaiser--Stebbins effect} \cite{gott85,kais-steb84}. 

The four statistical tests (MSD, SMD, skewness, kurtosis) have
been performed on patches of the maps (Fig.\ref{fig2}) produced by 
the simulation  in a way similar to the one in \cite{per-sim98}:
\begin{itemize} 
\item  First the values of the 4 statistical variables were computed
for a specific $30^{\circ}\times30^{\circ}$ patch obtained using 
the simulation code. 
\item A large number (1200) of Gaussianized maps were constructed
from this patch by randomizing the phases in Fourier space and
using a Gaussian spectrum with 0 mean and variance equal to the
measured spectrum of the map.
\item The  4 statistics were calculated for each of these
Gaussianized maps.
\item Using these results, the probability distribution for each
statistic was constructed.
\item Finally, the probability for obtaining the already calculated
(in step 1) values of the 4 statistics was found.
\end{itemize}

\begin{figure}[tbp]
 \centerline{\epsfxsize = 0.86\hsize \epsfbox{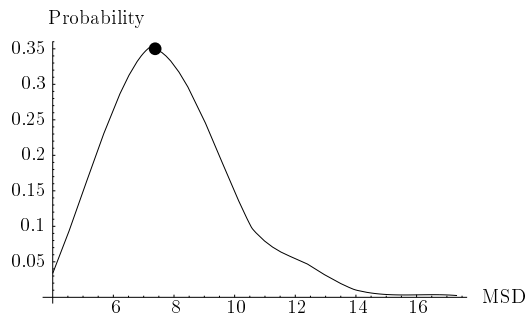}}
\caption[FIG.3]{Probability distribution of the MSD based on 1200 Gaussianized maps. 
The probability for obtaining the MSD value of the string--induced map 
(MSD=7.378, represented by the dot) is 0.35.}
\label{fig3}
\end{figure}

\begin{figure}[htbp]
 \centerline{\epsfxsize = 0.86\hsize \epsfbox{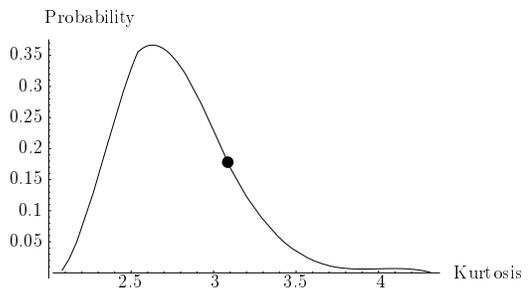}}
\caption[FIG.4]{Probability distribution of the kurtosis based on 1200 Gaussianized
maps. The probability for obtaining the kurtosis value of the string--induced map (kurtosis=3.08, represented by the dot) is 0.17.}
\label{fig4}
\end{figure}

The results are shown in Figs. \ref{fig3}, \ref{fig4}. Clearly,  
the applied statistics can not reveal non-Gaussianity in a 
statistically significant amount in the maps produced by the 
simulation. The values of the statistics considered for the string 
induced maps (black dots) are within $1\sigma$ in the probability 
distributions produced by the gaussianized maps. 

We have also extended the above results using right angles as 
partition lines instead of straight lines. However, the 
sensitivity of the statistics considered was not increased, 
rendering results similar to Figs. \ref{fig3}, \ref{fig4}. 

\section{Concluding Remarks}

String--induced maps on large angular scales are hard to
distinguish from maps with Gaussian fluctuations, even using
specially designed tests. We have suggested a technique for 
constructing string-induced temperature anisotropy maps which can 
be exploited in many ways. The maps constructed the way described
in section 2 are proved not to be dominated by late long strings.
On the contrary, small--scale structure features come into play
which, in a way, destructively interfere and ruin the large--scale
coherence which the MSD and SMD tests are proposed to optimally
detect. Thus, we offered a confirmation to the fact that 
string--induced temperature anisotropy maps are practically 
Gaussian on large scales. 

\section{Acknowledgements}

We would like to thank Levon Pogosian and Tanmay Vachaspati for 
their assistance in acquiring and better understanding the code 
they developed by modifying CMBFAST.

%\bibliographystyle{prsty}

%\bibliography{bibliog}

\end{document}